\newfam\msbfam
\batchmode\font\twelvemsb=msbm10 scaled\magstep1 \errorstopmode
\ifx\twelvemsb\nullfont\def\Bbb{\bf}
	\message{Blackboard bold not available. Replacing with boldface.}
\else	\catcode`\@=11
	\font\tenmsb=msbm10 \font\sevenmsb=msbm7 \font\fivemsb=msbm5
	\textfont\msbfam=\tenmsb
	\scriptfont\msbfam=\sevenmsb \scriptscriptfont\msbfam=\fivemsb
	\def\Bbb{\relax\expandafter\Bbb@}
	\def\Bbb@#1{{\Bbb@@{#1}}}
	\def\Bbb@@#1{\fam\msbfam\relax#1}
	\catcode`\@=\active
\fi
%
%
\font\eightrm=cmr8	\def\xrm{\eightrm}
\font\eightbf=cmbx8	\def\xbf{\eightbf}
\font\eightit=cmti8	\def\xit{\eightit}
\font\eighttt=cmtt8	\def\xtt{\eighttt}
\font\eightcp=cmcsc8
\font\nineit=cmti9
\font\ninerm=cmr9

%
%
%
\footline={\hfil}
\def\makefootline{\baselineskip=1.6cm\line{\the\footline}}
%
%
\newcount\refcount
\refcount=0
\newwrite\refwrite
\def\ref#1#2{\global\advance\refcount by 1
	\xdef#1{\the\refcount}
	\ifnum\the\refcount=1
	\immediate\openout\refwrite=\jobname.refs
	\fi
	\immediate\write\refwrite
		{\item{[\the\refcount]} #2\hfill\par\vskip-2pt}}
\def\refout{\catcode`\@=11
	\xrm\immediate\closeout\refwrite
	\vskip\baselineskip
	{\noindent\tenbf References}\hfill
						\vskip.5\baselineskip
	\parskip=.875\parskip 
	\baselineskip=.8\baselineskip
	\input\jobname.refs 
	\parskip=8\parskip \divide\parskip by 7
	\baselineskip=1.25\baselineskip 
	\catcode`\@=\active\rm}
%
%
\headline={\ifnum\pageno=1\hfill\else
{\eightcp Martin Cederwall: 
	``Problems with Duality in N=2 SYM''}
		\dotfill\folio\fi}
\def\makeheadline{\vbox to 0pt{\vss\noindent\the\headline\break
\hbox to\hsize{\hfill}}
	\vskip2\baselineskip}
%
%
%
\parskip=3.5pt plus .3pt minus .3pt
\baselineskip=12.5pt plus .1pt minus .05pt
\lineskip=.5pt plus .05pt minus .05pt
\lineskiplimit=.5pt
\abovedisplayskip=10pt plus 4pt minus 2pt
\belowdisplayskip=\abovedisplayskip
\hsize=12cm
\vsize=17.75cm
\hoffset=2cm
\voffset=3cm
%
%
\def\/{\over}
\def\*{\partial}

\def\Z{{\Bbb Z}}

\def\.{.\hskip-1pt }
\def\is{\!=\!}
\def\-{\!-\!}
\def\+{\!+\!}
\def\={\!=\!}

\def\f{{\lower1.5pt\hbox{$\scriptstyle f$}}}
\def\ff{{\lower1pt\hbox{$\scriptstyle f$}}}

\def\ie{{\it i.e.}\hskip-1pt}
\def\eg{{\it e.g.}\hskip-1pt}
%
%
%
%
%
\null\vskip-1cm
\vtop{\baselineskip=9pt
\hbox to\hsize{\hfill\xrm G\"oteborg-ITP-96-9}
\hbox to\hsize{\hfill\xtt hep-th/9606096}
\hbox to\hsize{\hfill\xrm June, 1996}
}

\vskip10\parskip
\centerline{\tenbf PROBLEMS WITH DUALITY}
 
\centerline{\tenbf IN N=2 SUPER--YANG--MILLS THEORY}
 
\vskip4\parskip
\centerline{\ninerm MARTIN CEDERWALL}

\vskip2\parskip
\centerline{\nineit Institute for Theoretical Physics}\vskip-4pt
\centerline{\nineit G\"oteborg University 
		and Chalmers University of Technology }\vskip-4pt
\centerline{\nineit S-412 96 G\"oteborg, Sweden}

\ref\CederwallHolm{M.~Cederwall and M.~Holm, {\xtt hep-th/9603134}.}
\ref\SeibergIV{N.~Seiberg and E.~Witten, 
	\xit Nucl.Phys. \xbf B426 \xrm(1994) 19;\hfill\break\indent
	\xrm erratum: ibid. \xbf B430 \xrm (1994) 485 ({\xtt hep-th/9407087}).}
\ref\SeibergIII{N.~Seiberg and E.~Witten, \xit Nucl.Phys. \xbf B431
	\xrm (1994) 484 ({\xtt hep-th/9408099}).}
\ref\Klemm{A.~Klemm, W.~Lerche and S.~Theisen, {\xtt hep-th/9505150}.}
\ref\Vafa{C.~Vafa and E.~Witten, \xit Nucl.Phys. \xbf B431 \xrm (1994) 3
	({\xtt hep-th/9408074}).}
\ref\Intriligator{K.~Intriligator and N.~Seiberg \xit Nucl.Phys. \xbf B431 
	\xrm (1994) 551 ({\xtt hep-th/9408155}).}
\ref\Montonen{C.~Montonen and D.~Olive, 
	\xit Phys.Lett. \xbf 72B \xrm (1977) 117.}
\ref\GNO{P.~Goddard, J.~Nuyts and D.~Olive, 
	\xit Nucl.Phys. \xbf B125 \xrm (1977) 1.}
\ref\Mandelstam{S.~Mandelstam, \xit Nucl.Phys. \xbf B213 \xrm (1983) 149.}
\ref\Brink{L.~Brink, O.~Lindgren and B.E.W.~Nilsson,
	\xit Phys.Lett. \xbf 123B \xrm (1983) 323.}
\ref\Sohnius{M.~Sohnius and P.~West, \xit Phys.Lett. \xbf 100B \xrm (1981) 45.}
\ref\Howe{P.S.~Howe, K.S.~Stelle and P.C.~West, \xit Phys.Lett. \xbf 124B
	\xrm (1983) 55.}
\ref\HoweII{P.S.~Howe, K.S.~Stelle and P.K.~Townsend,
	\xit Nucl.Phys. \xbf B214 \xrm (1983) 519.}
\ref\Piguet{O.~Piguet and K.~Sibold, 
	\xit Helv.Phys.Acta \xbf 63 \xrm (1990) 71.}
\ref\SeibergV{N.~Seiberg \xit Phys.Lett. \xbf 206B \xrm (1988) 75.}
\ref\Sen{A.~Sen, \xit Phys.Lett. \xbf B329 \xrm (1994) 217
	({\xtt hep-th/9402032}).}
\ref\Porrati{M.~Porrati, {\xtt hep-th/9505187}.\vfill\eject}
\ref\Cederwall{M.~Cederwall, G.~Ferretti, B.E.W.~Nilsson and P.~Salomonson,
	\hfill\break\indent\xit Mod.Phys.Lett. \xbf A11 \xrm (1996) 367
	({\xtt hep-th/9508124}).} 
\ref\Sethi{S.~Sethi, M.~Stern and E.~Zaslow,\xit Nucl.Phys. \xbf B457 
    \xrm (1995) 484 ({\xtt hep-th/9508117}).}
\ref\GauntlettIII{J.P.~Gauntlett and J.A.~Harvey, \xit Nucl.Phys. \xbf B463
    \xrm (1996) 287 ({\xtt hep-th/9508156}).}
\ref\GauntlettIV{J.P.~Gauntlett and D.A.~Lowe {\xtt hep-th/9601085}.}
\ref\Lee{K.~Lee, E.~Weinberg and P.~Yi, {\xtt hep-th/9601097}.}
\ref\Olive{D.~Olive and E.~Witten, \xit Phys.Lett. \xbf 78B \xrm (1978) 97.}

\vskip2\parskip
\noindent\hskip1cm\vtop{\hsize=10cm \baselineskip=9pt\xit \noindent
Actual calculations of monopole and dyon spectra have previously been  
performed in N=4 SYM and in N=2 SYM with gauge group SU(2),
and are in total agreement with duality conjectures for the finite theories. 
These calculations are extended to N=2 SYM with higher rank gauge 
groups, and it turns out that the SU(2) model with four fundamental 
hypermultiplet is an exception in that its soliton spectrum supports 
duality. This may be an indication that the other perturbatively finite N=2
theories have non-perturbative contributions to the 
$\scriptstyle\beta$-function.
This talk contains a short summary of recent results [\CederwallHolm].}

\vskip8\parskip\noindent
The last years have seen big progress in the understanding of 
non-perturbative aspects of quantum field theory and string theory.
In supersymmetric enough theories, some non-perturbative properties may
actually be calculated exactly 
[\SeibergIV,\SeibergIII,\Klemm,\Vafa,\Intriligator]. 
Intimately connected with the new 
techniques developed is the somewhat older concept of duality 
[\Montonen,\GNO]. Finite 
quantum field theories may, and seem to, exhibit a duality symmetry, which 
is the realization and extension of the old puzzling symmetry between 
electricity and magnetism in Maxwell theory. $\Z_2$ duality transforms the 
coupling constant of the theory to its inverse, and exchanges (non-abelian)
electric and magnetic charges. When the theta angle is included in the 
complex coupling constant, the group of duality transformations is 
$Sl(2;\Z)$, acting on the vector of electric and magnetic charge, and 
projectively on the coupling constant. Then also dyons, states carrying 
both electric and magnetic charge, are involved.
The relevant magnetically charged states 
in a model containing Yang--Mills and a Higgs field are the magnetic 
(multi-)monopoles and dyons. If a theory is finite, and the coupling constant 
does not run, the transformations can make sense, and there is the possibility 
that duality is an exact symmetry of the theory. Then there is no 
fundamental difference between elementary and solitonic excitations, 
just that we need to chose one element in the duality group, \ie\ a set of
``elementary'' excitations, in order to formulate the theory. A geometric 
understanding of this kind of duality, in the context of string theory 
called S-duality, is still lacking, and is probably needed for our 
understanding of what string theory ``really is''.

There is an infinity of perturbatively finite quantum field theories.
Of special interest among these are supersymmetric Yang--Mills theories with
at least two supersymmetries, \ie\ $N\is2,4$. All $N\is4$ super-Yang--Mills
theories are perturbatively finite 
[\Mandelstam,\Brink,\Sohnius,\Howe,\HoweII,\Piguet], 
and this property has been shown to
hold also non-perturbatively [\SeibergV]. 
The models with $N\is2$ are perturbatively
finite if the matter content is appropriate, \eg\ the $SU(N_c)$ models with
$N_f\is2N_c$ hypermultiplets in the fundamental representation. It is of 
course important to decide whether these theories are non-perturbatively
finite.

The questions of finiteness and duality are closely related. A 
non-running coupling constant is required for duality symmetry, and 
conversely, if there is a number of excitations that become massless as
the coupling constant goes to some specific value in a finite theory, 
these should play the role of elementary excitations in a perturbation 
expansion of a finite theory. Examining the spectrum of solitons in a 
theory may therefore give information both about duality and finiteness.

The procedure for finding (part of) the spectra of monopoles and dyons in 
the theories at hand is well established 
[\Sen,\Porrati,\Cederwall,\Sethi,\GauntlettIII,\GauntlettIV,\Lee]. 
Starting from a classical 
solution that is BPS-saturated, \ie\ has a certain relation between mass 
and magnetic charge and breaks half of the supersymmetry, one makes a 
low-energy approximation. 
Here, the $N\is2$ supersymmetry is crucial, in that the electric and 
magnetic charges enter the central charges in the supersymmetry algebra.
BPS-saturated states come in short multiplets, with half the size of a 
full multiplet (this is identical to the difference between massless and 
massive multiplets, only that ``$m\is0$'' is shifted in the presence of 
the central extensions). The number of states in a multiplet should not 
be quantum corrected and the BPS property is therefore valid also outside the 
low-energy approximation [\Olive]. 
Due to the presence of zero-modes, bosonic and 
fermionic, and the presence of a mass-gap, the low-energy approximation 
gets only a finite number of degrees of freedom, namely the moduli 
parameters of the relevant topological sector (specified by magnetic 
charge) and fermionic zero-modes. The task of finding BPS states is,
roughly speaking, 
reduced to that of finding ground states of a supersymmetric quantum 
mechanics model, which amounts to a cohomology problem.
The riemannian geometry of the monopole moduli spaces 
and the geometry of the index bundles of fermionic zero-modes over these 
moduli spaces are essentially determined from the form of the kinetic 
terms in the field theory lagrangian, and the geometric considerations 
can be pushed quite far on a purely formal basis. What restricts the 
actual calculation of ground states, is that it, at least this far, requires 
concrete knowledge of the moduli spaces. Only at the lowest values of the 
magnetic charge is the metric structure known explicitely, which 
restricts the calculations to these cases (note, however, the method of
[\Porrati], which seems to circumvent this statement).

These calculations have been performed for the $N\is4$ theories 
[\Sen,\Porrati,\GauntlettIV,\hfill\break\Lee], 
where the results are in perfect agreement with the predictions from 
duality. Also for the $N\is2$ theory with gauge group $SU(2)$ and four 
hypermultiplets in the fundamental representation, duality (as predicted 
in [\SeibergIII]), is in agreement with the obtained spectrum 
[\GauntlettIII,\Sethi]. We wanted to 
continue this investigation to include also the other perturbatively 
finite $N\is2$ theories. The expectation was to find spectra that would lend 
themselves to a 
direct interpretation in terms of a dual finite theory. 
For a number of reasons,
this did not happen.

It is quite easy to point at at least two problems with duality in $N\is2$ 
theories with gauge groups of rank higher that one. The electric charges 
of the elementary excitation lie on the weight lattice of the gauge 
group, and the magnetic charges on the dual to this lattice (more precisely,
on the dual to the weight lattice of the universal cover of the group), 
the coroot lattice, which for simply laced groups coincide with the root lattice.
These two lattices are in general different, also modulo a rescaling
(for $SU(4)$, they are the {\it fcc} and {\it bcc} lattices). 
Note that this only is an argument against $\Z_2$ symmetry of 
electric--magnetic interchange in the cases where the lattices differ, 
but in a finite theory this would also mean that one is not expecting to 
find any BPS-saturated states with purely magnetic charge, which we 
actually do. Also, there is a serious problem already with the classical 
moduli spaces, namely that only some sectors of the coroot lattice are 
allowed as magnetic charges. These are the coroots that in a suitable 
basis are sums of the simple coroots with only positive or only negative 
coefficients. The other coroots would correspond to superpositions of 
monopoles and anti-monopoles, \ie\ of selfdual and anti-selfdual field 
configarations. Such a configuaration can not be BPS-saturated. In the 
$N\is 4$ case, where only the trivial conjugacy class of the weight 
lattice is populated with elementary excitations, this presents no 
problems. All the magnetic charges align with the magnetic ones, and the 
forbidden sectors are avoided. For $N\is2$ theories with gauge groups of 
rank higher than one this is a problem. 

Of course, indirect arguments like these, how convincing they might sound,
can turn out to be wrong or irrelevant for reasons difficult to predict.
Therefore, it is important to perform the actual calcualtion of the spectra
in order to examine if there is some obvious sign of duality in some 
parts of them. We did not find any such signs. Some of the spectra for 
gauge group $SU(3)$ are presented diagrammatically in [\CederwallHolm].

What are the interpretations of these results?
One must bear in mind that the non-perturbative finiteness of all these 
$N\is2$ models has not been proven, and the results we have obtained 
would become very natural if there were instanton contributions to the 
$\beta$-function (with the exception of the finite $SU(2)$ theory).
We do not think our evidence is strong enough for a claim of this type,
but we feel that the possibility must be considered, even if it is 
against the common lore.
Presumably, if the perturbatively finite models discussed here actually exhibit
non-perturbative divergencies, they should manifest themselves as diverging
sums over instanton number, since the contribution from a sector 
with given instanton number should be perturbatively finite, and the 
integrals over instanton moduli converge in the presence of a Higgs 
expectation value.

For concrete calculations, formulas and presentation of the spectra, 
I refer to [\CederwallHolm], which also contains a more exhaustive list 
of references.

\vskip\baselineskip
\noindent{\tenbf Acknowledgement}

\noindent The work summarized here was done in collaboration with Magnus Holm. 
I would also like to thank the organizers of the Second International Sakharov 
Conference on Physics for their great hospitality and all the help they 
provided during this exciting week.

\refout
\end